\input {epsf}

\documentclass [12pt] {article}

\usepackage [cp1251] {inputenc}

\usepackage {graphicx}
\input {epsf}

\begin {document}
\author {Kozlov G.G.}
\title {The Correlated Lloyd model: exact solution}
\maketitle
\begin {abstract}
Exactly solvable model of disordered  system representing the generalized Lloyd model 
with correlated random potential  is described. 
It is shown, that for  the model under consideration, the averaged Green's function does 
not depend on random potential correlation radius   and, similarly to the classical Lloyd 
model, has the form of    Green's function of a crystal system, with energy argument supplied  
by  an imaginary part which depends on  degree of  disorder.
\end {abstract}

\section * {Introduction}

 The number of annually published papers devoted to the models of disordered  systems,  has noticeably  increased during the last decade,
 The "center of gravity'' of up to date research is shifting 
towards {\it correlated} disordered systems, exploration of which 
 becomes more and more popular \cite {Izr,Tit,Deych,Der,Croy,Mal3,Fran,Dunlap,Koz1,Koz0}.
  In the historically first models of random systems, as a rule, the 
$\delta$-correlated random potential was employed and relatively small attention 
was paid to the  influence of correlations.
The recent research have shown, that  correlations can lead to a considerable (sometimes, qualitative)
modification of energy structure and localisation properties  of disordered systems.

   The exactly solvable models play an important role in an arbitrary field of theoretical physics. Exactly solvable model allows one to accumulate the qualitative information
  related to the appropriate class of models with a high degree of reliability, to test used approximations, to specify a direction of further research, etc.
  The most known and important exactly solvable models in physics of disordered systems  are  Dyson model  \cite {Dyson} and
  Lloyd model \cite {Lloyd}. Both these models relate to the uncorrelated disorder.
 In this paper we propose a  generalized one-dimensional Lloyd model,  with  {\it site energies being not an  
independent random variables} and describe  the exact calculation  of the averaged Green's function,
 and  show, that it does not depend on 
the  model's parameter which  plays the role of correlation radius.

     The paper is organised  as follows. 
In the first section we consider the classical Lloyd model, formulate definitions,  
necessary  for further analysis  and  obtain some auxiliary results.  
 In the second section we introduce 
the correlated disorder, for which 
 the exact evaluation of the averaged Green's function is carried out in the third section. 
 
\section {Lloyd model}

 Matrix $ {\bf H} $ of the Hamiltonian of  classical one-dimensional Lloyd model  has the following elements:
\begin {equation}
 H_{rr'}=\delta_{rr'}\varepsilon_r+w(r-r')\hskip10mm r, r ' =1,2,3..., N
\label {1a}
\end {equation}
where  function $w (r) $ (with $w (0) =0 )$ is specified, 
 and diagonal elements (the site energies) $ \varepsilon_r $ 
 are independent, similarly  distributed random variables with  Cauchy distribution   function 
\begin {equation}
P (\varepsilon) = {1\over \pi} {\Delta\over \Delta^2 +\varepsilon^2}
\label {Coch}
\end {equation}
  Parameter $ \Delta $ characterizes a degree of disorder
 -- at $ \Delta\rightarrow 0 $ the Hamiltonian (\ref {1a}) corresponds
to ordered (crystalline) system and can be diagonalised in the representation of plane
waves. A set of numbered site energies is frequently referred to as a random  potential.

\begin {figure}
 \begin {center}
\includegraphics [width=10cm] {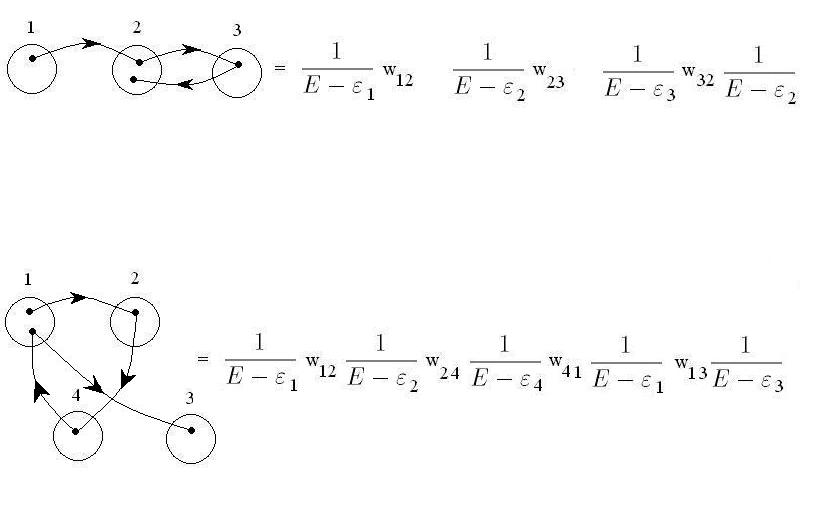}
\caption {Examples of the diagrams}
 \label {fig1}
 \end {center}
\end {figure}

 In his famous work \cite {Lloyd} Lloyd has managed to calculate exactly the  averaged
Green's function  $ \langle {\bf G} (\Omega) \rangle =\langle [\Omega- {\bf H}] ^ {-1} \rangle $,
( $ \Omega\equiv E +\imath 0 $) for a model system (\ref {1a}), (\ref {Coch}),
 which (Green function) allows one to calculate density of states and  spectrum of  linear susceptibility.
In this section we  reproduce Lloyd's result  by means of a dyagram  technique,
similar to that offered in \cite {GAF} (see also \cite {Koz2}). 
Let's introduce a matrix $ {\bf W} $ with elements $W _ {rr '} \equiv w (r-r ') $. Then
the Green's function matrix  $ {\bf G} $ can be written as the following series \cite {Lif}
\begin {equation}
G _ {rr '} (\Omega) = {\delta _ {rr '} \over \Omega-\varepsilon_r} + {1\over
\Omega-\varepsilon_r} W _ {rr '} {1\over \Omega-\varepsilon_r '} + \sum _ {r'' }
{ 1\over \Omega-\varepsilon_r} W _ {rr ''} {1\over \Omega-\varepsilon_r ''} W _ {r ''r'} {1\over \Omega-\varepsilon_r} +...
\label {3a}
\end {equation}
Let's denote  matrix element $W _ {r, r '} $ by an arrow, directed from site $r$ to site $r'$ and 
denote factor $ [\Omega-\varepsilon_r] ^ {-1} $, related  to the site $r$, 
by a bold dot,  placed inside the appropriate site.  The examples  are presenteded at fig.\ref {fig1}.
Then  one can write the following expression for the matrix element $G _ {rr '} (\Omega) $ of the Green's function
\begin {equation}
G _ {rr '} (\Omega) = \hbox {the sum of all diagrams pairing sites $r $ and $r ' $}
\label {4a}
\end {equation}
To calculate the sought  averaged Green's function one should integrate
(\ref {4a}) with a joint distribution function of site energies $ \varepsilon_1..., \varepsilon_N $. In the
  case of  uncorrelated disorder this function can be represented as  a  product:
 \begin {equation}
  \rho _ {nc} (z_1, z_2..., z_N) = \prod _ {i=1} ^N P (z_i),
  \label {5a}
  \end {equation}
   For Lloyd model under consideration the function
 $P (z) $ has the form (\ref {Coch}). 
Averaging of an arbitrary diagram $D$ in  expansion (\ref {4a}), is  reduced to  
 averaging of  factor $f_D $ defined  as:
 \begin {equation}
 f_D\equiv \bigg ({1\over \Omega -\varepsilon _ {n_1}} \bigg) ^ {g_1}
 \bigg ({1\over \Omega-\varepsilon _ {n_2}} \bigg) ^ {g_2} ...\bigg ({1\over \Omega -\varepsilon _ {n_q}} \bigg) ^ {g_q}
\label {6a}
\end {equation}
where $n_1, n_2..., n_q $ --  are the numbers of sites, the diagram  $D $ passed through, and $g_i$ is the
 number of diagram's passages through the site 
  $n_i, i=1,2,...,q$ (number of bold dots inside the site $n_i$). 
 For example, for  upper diagram at  fig.\ref {fig1} we have:
 $ n_1=1, n_2=2, n_3=3 $ and $g_1=1, g_2=2, g_3=1 $. The key point for Lloyd's 
 solution is the following rule for calculating of  relevant integrals  
 \begin {equation}
     {1\over \pi} \int {\Delta dz\over \Delta^2+z^2} \bigg ({1\over E +\imath 0 -z} \bigg) ^n =
     \bigg ({1\over E +\imath \Delta} \bigg) ^n
\label {7a}
\end {equation}
     Taking advantage of  this relationship and implying  the site 
energies $ \varepsilon _ {n_i}, i=1,2..., q $ to be mutually independent, we obtain
the following expression for the averaged factor $f_D $:
     \begin {equation}
     \langle f_D\rangle =
     \int \prod _ {i=1} ^Ndz_i P (z_i)
     \bigg ({1\over \Omega -z _ {n_1}} \bigg) ^ {g_1}
      \bigg ({1\over \Omega-z _ {n_2}} \bigg) ^ {g_2} ...\bigg ({1\over \Omega -z _ {n_q}} \bigg) ^ {g_q}
     =
\label {8a}
 \end {equation}
     $$
     = \bigg ({1\over \Omega + \imath\Delta} \bigg) ^ {g_1}
      \bigg ({1\over \Omega +\imath\Delta} \bigg) ^ {g_2} ...\bigg ({1\over \Omega + \imath\Delta} \bigg) ^ {g_q}
 $$
which is  coincides with that for a diagram $D $ of the Green's function of the  {\it  ordered} system with all site 
energies  being zero, and with  energy argument being replaced as: $\Omega\rightarrow\Omega +\imath\Delta $. 
 The above calculation  is valid for any diagram in  expansion (\ref {4a}) and we come to the Lloyd's result:
  {\it the averaged Green function of disordered system described by the Hamiltonian (\ref {1a}) 
with the Cauchy uncorrelated disorder  (\ref {Coch}) is equal to a Green's function $ {\bf G} ^ {od} $ of the 
 ordered system with  zero site energies, and in which
the  replacement $\Omega\rightarrow\Omega+\imath\Delta$ of energy argument
    is made}: 
  \begin {equation}
  \langle {\bf G} (\Omega) \rangle = {\bf G} ^ {od} (\Omega +\imath\Delta)
  \label {9a}
  \end {equation}
     The explicit expression for   a Green's function matrix of ordered (and cyclic)
      system $ {\bf G} ^ {od} $ can be obtained by  using the fact  that
      eigenvectors  of the Hamiltonian (\ref {1a}) at $ \varepsilon_r=0\hskip3mm (r=1..., N) $ 
are plane waves \cite{Lif}. For the case of  one-dimensional system this matrix is:
     \begin {equation}
     G ^ {od} _ {rr '} (\Omega) = {1\over 2\pi} \int _ {-\pi} ^ \pi {e ^ {\imath {(r-r ') q}} \over \Omega-J _ {q}} \hskip1mm dq, \hskip5mm
     J _ {q} = \sum _ {r} w (r) e ^ {-\imath {q r}}
\label {10a}
    \end {equation}

     In the next section we  describe the simple correlated discrete random process  
 $\varepsilon_1,..\varepsilon_i,...,\varepsilon_N$, for which the  total joint  distribution function $ \rho (z_1, z_2... z_N) $ (which can not be reduced to the product like (\ref {5a})) can be calculated in the close form.
    In the final, third, section we  show, that for the random Hamiltonian  (\ref {1a}),
 with site energies represented by   such random process, the result (\ref {9a}) holds.

       \section {Correlated disorder}

We obtain the correlated sequence of site energies $ \varepsilon_r, r=1..., N $
  by the following procedure {\it of smoothing}
  \footnote {the similar mechanism of correlation was employed in \cite {Koz1} for calculation
 of a degree of localization of states in the correlated system.} \cite {Prob, Prob1}.
  Let's introduce  a set  of independent random values
   $ \xi_i, i =-\infty..., -1,0,1..., + \infty $, each of which  has the specified distribution function
    $P (\xi) $ (at this stage of calculation it can differ from (\ref{Coch})), independent on  $i $.
Now obtain the site energies $ \varepsilon_n $ as a realisation of the following discrete random process:
      \begin {equation}
          \varepsilon_n = (1-e ^ {-\alpha}) \sum _ {m\le n} e ^ {\alpha (m-n)} \xi_m, \hskip10mm \alpha > 0
          \hskip5mm n=1..., N
          \label {11a}
          \end {equation}
The values $ \varepsilon_n $ will be correlated with a correlation radius  
$R=1/\alpha $. The relevant correlation function $ \langle \varepsilon_n\varepsilon _ {n '} \rangle $ has the form 
\begin {equation}
\langle \varepsilon_n\varepsilon_{n'}\rangle=\langle \xi^2\rangle\bigg ({1-e ^ {-\alpha} \over
1+e ^ {-\alpha}} \bigg) \hskip2mm e ^ {-\alpha|n-n ' |} = \langle \xi^2\rangle\bigg ({1-\beta\over
1 +\beta} \bigg) \hskip2mm e ^ {-\alpha|n-n ' |}, \hskip5mm\beta\equiv e ^ {-\alpha}, \hskip2mm \beta\in [0,1]
\label {12a}
\end {equation}
From definition (\ref {11a}) it is easy to obtain the following (important for the further) relationship:
\begin {equation}
 \varepsilon_{n+1}=\beta\varepsilon_n+(1-\beta)\xi_{n+1} 
\label {3}
\end {equation}
The correlation function (\ref {12a}) is meaningful only for  the case of 
finite second moment of the function $P (\xi) $ and is not exist if $P (\xi) $ has the form 
 (\ref {Coch}). Nevertheless, even in this case
it is not correct to consider a sequence (\ref {11a}) as uncorrelated, because, as we shall see below, its joint distribution function can not be presented in the form  (\ref {5a}).
   At last, we note, that the correlated sequence (\ref {11a}) is causal -- i.e.
  $ \varepsilon_n $ depends only on those $ \xi_m $, for which $m\le n $.

\subsection {Site energy  distribution function  for the case of correlated process (\ref{11a})}

The distribution function of an arbitrary  site energy $ \varepsilon_n $
(we shall denote it $ \sigma (\varepsilon) $) does not depend on  its number
$ n $ and we   calculate it for $n=0 $ \cite {Prob, Prob1}. As a starting point we use  
the following general expressions for the sought function
 $ \sigma (\varepsilon) $ and the relevant characteristic function $ \tilde\sigma (t) $:
\begin {equation}
\sigma (\varepsilon) = \bigg\langle \delta\bigg(\varepsilon-[1-\beta]\sum_{m=0}^\infty 
\beta^m\xi_m\bigg) \bigg\rangle\equiv\int e ^ {\imath \varepsilon t} \tilde\sigma (t)
\label {10}
\end {equation}
\begin {equation}
\tilde\sigma (t) = {1\over 2\pi} \bigg\langle\exp-\imath t\bigg ([1-\beta] \sum _ {m=0} ^ \infty
\beta^m\xi_m\bigg) \bigg\rangle = {1\over 2\pi} \prod _ {m=0} ^ \infty \int d\xi P (\xi) e ^ {-\imath t [1-\beta] \beta^m\xi}
\label {11}
\end {equation}
Here angular brackets correspond to an averaging over independent auxiliar variables $ \xi_m $.
 Denoting the Fourier transformation of the   function $P (\xi) $ as $ \tilde P (t) $:
$ \tilde P (t) = \int P (\xi) e ^ {-\imath t\xi} d\xi $,
 we obtain the following formula for $ \tilde\sigma (t) $:
\begin {equation}
\tilde\sigma (t) = {1\over 2\pi} \prod _ {m=0} ^ \infty \tilde P\bigg (t [1-\beta] \beta^m\bigg)
\label {13}
\end {equation}
If $P (\xi) $ is the Cauchy function (\ref {Coch}), then
 \begin {equation}
 \tilde P (t) =e ^ {- | t\Delta |}
 \hskip5mm\Rightarrow\hskip5mm
 \tilde\sigma (t) = {1\over 2\pi} \exp\bigg(-{|t|[1-\beta]\Delta\sum_{m=0}^\infty\beta^m} \bigg) =
{ 1\over 2\pi} e ^ {- | t |\Delta}
\label {14}
 \end {equation}
 Therefore 
  the  distribution function of site energies in our case also has the form  of  Cauchy (\ref {Coch}) function
\begin {equation}
 \sigma(\varepsilon)=P(\varepsilon)={1\over \pi} {\Delta\over \Delta^2 +\varepsilon^2}
\label {16}
\end {equation}

\subsection {Total distribution function of a random process (\ref {11a})}

 The discrete correlated random process $ \varepsilon_n $ (\ref {11a})
 is completely determined
 by  joint distribution function of all site energies $ \rho (z_1, z_2..., z_N) $,
 which can be calculated analytically.
  For this purpose we shall introduce the 
 joint probability distribution functions of the first $M $ ($ 0 < M < N $) site energies $ \rho_M (z_1, z_2..., z_M) $.
  So, the value  $ \rho_M (z_1, z_2..., z_M) dz_1... dz_M $
gives a probability that $ \varepsilon_i\in [z_i, z_i+dz_i], i=1..., M $.
 Using equation (\ref {3}) one can obtain  the recurent  relationship expressing $ \rho _ {M+1} $ through $ \rho_M $:
   \begin {equation}
    \rho_{M+1}(z_1,z_2,...,z_{M+1})=\bigg\langle \delta(z_1-\varepsilon_1)...\delta(z_M-\varepsilon_M)\delta(z_{M+1}-\varepsilon_{M+1}) \bigg\rangle =
   \end {equation}
 $$
  = \bigg\langle \delta(z_1-\varepsilon_1)...\delta(z_M-\varepsilon_M)\delta(z_{M+1}-\beta\varepsilon_{M}-[1-\beta]\xi_{M+1}) \bigg\rangle =
 $$
 $$
 = \int d\xi dy_1... dy_M \hskip2mm \rho _ {M} (y_1..., y_M) P (\xi) \hskip1mm \delta(z_1-y_1)...\delta(z_M-y_M)\delta(z_{M+1}-\beta y _ {M} - [1-\beta] \xi) =
 $$
 $$
 = {1 \over 1-\beta} \hskip1mm \rho _ {M} (z_1..., z_M) \hskip1mm P\bigg ({z _ {M+1} -\beta z_M\over 1-\beta} \bigg)
 $$
 Sequentially applying this relationship and taking into account, that $ \rho_1 (z) = \sigma (z)$,
 we obtain the following expression for the function $ \rho_M (z_1..., z_M) $
  \begin {equation}
   \rho _ {M} (z_1, z_2..., z _ {M}) = {1 \over [1-\beta] ^ {M-1}} \hskip1mm
   \overbrace {
   P\bigg ({z _ {M} -\beta z _ {M-1} \over 1-\beta} \bigg)
   P\bigg ({z _ {M-1} -\beta z _ {M-2} \over 1-\beta} \bigg)... P\bigg ({z _ {2} -\beta z _ {1} \over 1-\beta} \bigg)} ^ {M-1}
\hskip1mm \sigma (z_1)
   \label {34}
   \end {equation}
  In  our case
 $ \sigma (z) =P (z) $, where $P (z) $ is defined by  formula (\ref {Coch}). Supposing
     $ M=N $, we obtain the following final expression for the total joint  distribution function
 of  random process (\ref {11a})
   \begin {equation}
     \rho (z_1, z_2..., z _ {N}) =
     \overbrace {
     P\bigg ({z _ {N} -\beta z _ {N-1} \over 1-\beta} \bigg)
     P\bigg ({z _ {N-1} -\beta z _ {N-2} \over 1-\beta} \bigg)...
P\bigg ({z _ {2} -\beta z _ {1} \over 1-\beta} \bigg)} ^ {N-1} \hskip1mm
{ P (z_1) \over [1-\beta] ^ {N-1}}
     \label {21}
     \end {equation}
     $$
     P (z) = {1\over \pi} {\Delta\over \Delta^2+z^2}
     $$

\section {Correlated Lloyd model}

Our task  now is to calculate  the averaged Green's function of disordered model 
 described by the Hamiltonian $ (\ref {1a}) $,
 with site energies representing the realisation of correlated random process (\ref {11a}).
 Let's consider, as well as in the first section,  an arbitrary diagram $D $, passing 
 through the sites $n_1,n_2,...,n_q$, whose numbers without loss of generality we can
  consider to be arranged in ascending order: $1\le n_1 < n_2 <... < n_q\le N $.
 The average value $ \langle f_D\rangle $ of  factor (\ref {6a}) of the  diagram under consideration is
 now defined  by a formula, differed from (\ref {8a}):
          \begin {equation}
    \langle f_D\rangle = \int dz _ {1} dz _ {2}... dz _ {N} \bigg ({1\over E +\imath\delta -z _ {n_1}} \bigg) ^ {g_1}
   \bigg ({1\over E +\imath\delta -z _ {n_2}} \bigg) ^ {g_2} ...\bigg ({1\over E +\imath\delta -z _ {n_q}} \bigg) ^ {g_q} \rho (z_1, z_2...., z_N)
    \label {22}
     \end {equation}
      where function $ \rho (z_1, z_2...., z_N) $
 for  our correlated
 Lloyd model   has the form (\ref {21}).
 As $ \delta > 0 $, the factors in big  brackets
in the relationship (\ref {22}), considered
 as a functions of complex $z _ {n_1}..., z _ {n_q} $, {\it have no
 singularities in a lower half-plane of complex} $z _ {n_1}..., z _ {n_q} $. This
 allows us to use  formula (\ref {7a})
for evaluation of integrals figuring in the formula (\ref {22})  as follows.

First of all let's integrate the relationship (\ref{22})
 over all $z_i $  with $n_q < i\le   N$. 
After that the function (\ref {21}), entering this relationship, will lose its first
  $N-n_q $ factors of a type
$P\bigg ({z _ {N} -\beta z _ {N-1} \over 1-\beta} \bigg) P\bigg ({z _ {N-1} -\beta z _ {N-2} \over 1-\beta} \bigg)... $
and denominator $ [1-\beta] ^ {N-1} $ will be replaced by $ [1-\beta] ^ {n_q-1} $.
  Then perform the integration over $z _ {n_q} $, which affects only
  the  Lorentzian
 $P\bigg ({z _ {n_q} -\beta z _ {n_q-1} \over 1-\beta} \bigg) $, entering (\ref {21}),  and
  can be carried out with the help of formula (\ref {7a}). This formula shows that  
 mentioned integration corresponds to multiplication by the factor $1-\beta $ and
  to replacement $z _ {n_q} \rightarrow \beta z _ {n_q-1} -\imath\Delta (1-\beta) $
in the factor  
$ \bigg ({1\over E +\imath\delta -z _ {n_q}} \bigg) ^ {g_q} $,
entering the integrand in (\ref {22}).
  Thus the pole of this function (with respect to  $z _ {n_q-1} $) 
  will still be placed  in the upper half plane.
  This allows to perform the next integration
 over  $z _ {n_q-1} $ in the same manner. The only function entering  joint probability density (\ref {21})
   depending on this argument is $P\bigg ({z _ {n_q-1} -\beta z _ {n_q-2} \over 1-\beta} \bigg) $.
As well as in the previous case, the integration over $z _ {n_q-1} $
 corresponds to multiplication by the factor $1-\beta $ and replacement
$ z _ {n_q-1} \rightarrow \beta z _ {n_q-2} -\imath\Delta (1-\beta) $ in factor
      $ \bigg ({1\over E +\imath\delta -\beta z _ {n_q-1} + \imath\Delta (1-\beta)} \bigg) ^ {g_q}$,
  which arose as a result of the previous integration, etc.
  Hence, every new integration over $z_i $
  with lower and lower number $i $ corresponds to multiplication by $1-\beta $
(i.e. erasing of such a factor in the denominator of  expression (\ref {21})) and to replacement
$ z _ {i} \rightarrow \beta z _ {i-1} -\imath\Delta (1-\beta) $
  in the last term.
 
 When the number $i $ of a variable of  integration will decrease down to  $i=n _ {q-1} $, 
the further integrations can be performed  similarly (i.e. by making replacements of arguments),
 but  now the   above replacements  should be performed  in the factor
$ \bigg ({1\over E +\imath\delta -z _ {n _ {q-1}}} \bigg) ^ {g _ {q-1}} $ as well.

 Thus, we come to the conclusion, that  integration over all variables in (\ref {22})
 corresponds  to sequential replacements of symbols 
  $z _ {n_q}..., z _ {n_2}, z _ {n_1} $
 figuring in (\ref {22})   in accordance with  the above  rules. 
For example, the appropriate replacements for  $z _ {n_q} $ have the form:
 \footnote {As $n_q $ is the major of sites numbers $n_i $
   of the diagram under consideration,
 the replacements of remaining symbols are included in  the sequence, presented   below.}
\begin {equation}
 z _ {n_q} = \beta z _ {n_q-1} -\imath\Delta (1-\beta), \hbox {where}
\label {23}
\end {equation}
$$
z _ {n_q-1} = \beta z _ {n_q-2} -\imath\Delta (1-\beta), \hbox {where}
$$
$$
z _ {n_q-2} = \beta z _ {n_q-3} -\imath\Delta (1-\beta), \hbox {where...}
$$
$$
....
$$
$$
z _ {3} = \beta z _ {2} -\imath\Delta (1-\beta), \hbox {where...}
$$
$$
z _ {2} = \beta z _ {1} -\imath\Delta (1-\beta).
$$
The last integration over $z_1 $ corresponds to replacement $z_1\rightarrow -\imath\Delta $,
as this integration is performed with function $P (z_1) $ (see (\ref {21})).
It is easy to see, that if $z_1 = -\imath\Delta $, the chain of replacements (\ref {23}) 
 simplifies  and corresponds
  to the  replacement $z _ {n_i} = -\imath\Delta, i=1..., n_q $. Thus, the average (\ref {22}) corresponds 
  to the replacement
  of all symbols $z _ {n_i}, i=1.., q $ with $ -\imath\Delta $, as well as in the 
 case of uncorrelated Lloyd model (\ref {8a}), and
 we come to the conclusion, that
  {\it the averaged Green's function of the correlated Lloyd model 
  with site energies in the form (\ref {11a}) does not depend on the correlation radius   $R =-1/\ln\beta $
  and appears  to be the same,  as at lack of correlations}, i.e. is defined
   by  formula (\ref {9a}).

\begin {figure}
 \begin {center}
\includegraphics [width=10cm] {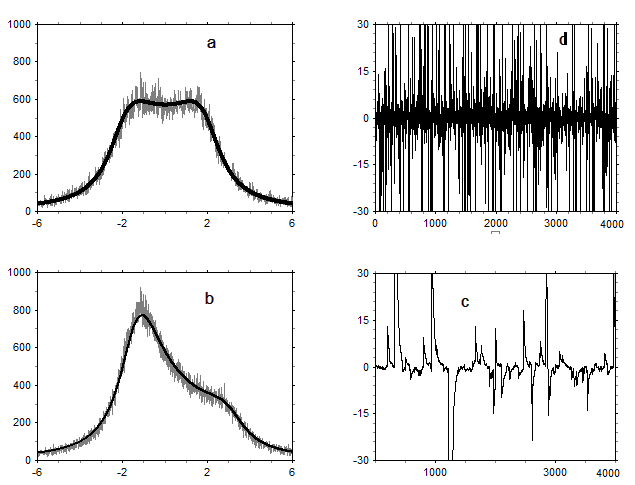}
\caption {The correlated Lloyd model  at various radiuses of correlation $R $ and various 
types of function $w (r) $, describing the nondiagonal elements of the Hamiltonian (\ref {1a}).
Panels (a) and (b) -- density of states of the Hamiltonian (\ref {1a}) 
for $w (r) = \delta _ {1, r} + \delta _ {-1, r} $ (a), and 
for $w (r) =v_0\exp - |r/R_1 | $ ($R_1=1, v_0 =\exp [1/R_1] $) (b). The noisy dependences 
on boards (a) and (b)  are obtained by  numerical diagonalization of the Hamiltonian 
(\ref {1a}), smooth curves -- calculation by formulas (\ref {25}) and (\ref {27}) respectively. 
 The densities of states shown on panels  (a) and (b) are obtained for  correlation 
radiuses  $R=0.1 $ and $R=30 $ respectively. Boards (d) and (c) -- realisations of random  
potentials for  correlation radiuses  $R=0.1 $ and $R=30 $ respectively. 
 On an abscissa axis the site number  $n $, on an axis of 
 ordinates -- $ \varepsilon_n $ (\ref {11a}) is postponed. In all cases $ \Delta=1 $. 
 The size of random matrixes for  numerical calculations $N=4000$.}
 \label {fig2}
 \end {center}
\end {figure}

Lack of dependence of  the averaged Green's function on  correlation radius  $R =-1/\ln\beta $
 reveals an original scale invariance of the considered correlated Lloyd model,
 because the spatial dependence (dependence on the site number) of random potential 
 $ \varepsilon_n $ (\ref {11a}) appears to be essentially not the same for various 
 $R $ (fig. \ref {fig2} (d, c)).

Let's illustrate the obtained result by examples, 
when  Green's function $ {\bf G} ^ {od} (\Omega) $ (\ref {10a}) can be calculated analytically. 
First (well known)  example of this kind  is the case of 
tight-binding Hamiltonian with 
appropriate matrix $H^{tb}_{rr'}=\delta_{r,r'+1}+\delta_{r,r'-1}$ and diagonal elements of Green's function
   defined as $G^{od}_{nn}(\Omega)=[\Omega^2-4]^{-1/2}$ \cite{Lif}. According to the results obtained above,
 the average density of states $ \rho ^ {tb} _ \Delta (E) = -\pi ^ {-1} $ Im Sp $ \langle {\bf G} \rangle $
 of the random matrix (\ref {1a}) with site energies
 $ \varepsilon_r $ in the form (\ref {11a}) and with $w (r) = \delta _ {r, 1} + \delta _ {r, -1} $,
does not depend on  correlation radius  $R =-1/\ln\beta $ and can be calculated as:
 \begin {equation}
\rho ^ {tb} _ \Delta (E) =
- {N\over \pi} \hbox {Im} {1\over \sqrt {(E +\imath\Delta) ^2-4}}
\label {25}
\end {equation}

The second (less known) example is the case, when the matrix of ordered Hamiltonian has the form:
$H ^ {ex} _ {rr '} =v_0\exp {- |r-r ' | /R_0} $. In this case the Green's function matrix   is 
 defined by the relationship \cite{Koz3}:
  \begin {equation}
   \Gamma _ {rr '} (\Omega) =A\exp (- |r-r ' | \eta) + \hskip2pt {\delta _ {rr '} \over \Omega}
\label {26}
 \end {equation}
 where
$$
 A\equiv {V\over (\Omega-V) \hskip2pt \Omega \hskip2pt\sqrt {1- T^2}},
 \hskip5mm
 V\equiv v_0\hbox {th} \bigg ({1\over R_0} \bigg), \hskip10pt
 {1\over T} \equiv {V-\Omega\over \Omega} \hbox {ch} \bigg ({1\over R_0} \bigg), \hskip10pt
 \hbox {ch} \hskip1pt \eta\equiv
 \bigg | {1\over T} \bigg |
 $$
    and the density of states of  random matrix (\ref {1a}) with $w (r) =v_0\exp {- |r-r ' | /R_0} $ and
 $ \varepsilon_r $ (\ref {11a}) is defined by the expression:
\footnote {The occurrence of shift $v_0 $ in this formula  is 
 the consequence of the fact  that  Green's function (\ref {26}) is  
obtained for  matrix $ {\bf H} ^ {ex} $ whose diagonal elements are nonzero and  equal to $v_0 $.}
\begin {equation}
\rho ^ {ex} _ \Delta (E) =
- {N\over \pi} \hbox {Im } \Gamma _ {00} (E +\imath\Delta+v_0)
\label {27}
\end {equation}
 For these two cases the numerical diagonalization of random matrixes  (\ref {1a}) for various   
correlation radiuses  $R $
 has shown, that the density of states is described by  formulas (\ref {25}) and (\ref {27}) 
(to within noise) and 
really does not depend on correlation radius $R$, 
despite the fact, that random potential (\ref {11a}) reveals strong dependence on $R$ (fig. 2 (d,c)).

   \section * {Conclusion}

    The exact  calculation of  averaged Green's function for the correlated Lloyd model  
is presented. It is shown, that for a considered type of correlated random potential the averaged 
Green's function does not depend on parameter of a random potential playing a role of  
correlation radius. The  obtained result is verified  by numerical calculations.

Finally we want to emphasize, that the above result is valid for an arbitrary 
dimensionality of  the lattice, if indexing of sites and their random energies  satisfying the relationship  (\ref {11a}).

\end {document}